\documentclass{article}
\usepackage{graphicx} 
\usepackage{xcolor}
\usepackage{cite}
\usepackage{booktabs, multirow, enumitem} 
\usepackage{bm}
\usepackage{amsmath}
\usepackage{bbding}
\usepackage{url}

\title{Health Misinformation Detection in Web Content via Web2Vec: A Structural-, Content-based, and Context-aware Approach based on Web2Vec}
\author{Rishabh Upadhyay, Gabriella Pasi, and Marco Viviani}
\date{Preprint Version: June 2021}

\begin{document}

\maketitle

\begin{abstract}
In recent years, we have witnessed the proliferation of large amounts of online content generated directly by users with virtually no form of external control, leading to the possible spread of misinformation. The search for effective solutions to this problem is still ongoing, and covers different areas of application, from opinion spam to fake news detection. A more recently investigated scenario, despite the serious risks that incurring disinformation could entail, is that of the online dissemination of health information. 
Early approaches in this area focused primarily on user-based studies applied to Web page content. More recently, automated approaches have been developed for both Web pages and social media content, particularly with the advent of the COVID-19 pandemic. These approaches are primarily based on handcrafted features extracted from online content in association with Machine Learning. In this scenario, we focus on Web page content, where there is still room for research to study structural-, content- and context-based features to assess the credibility of Web pages.
Therefore, this work aims to study the effectiveness of such features in association with a deep learning model, starting from an embedded representation of Web pages that has been recently proposed in the context of phishing Web page detection, i.e., Web2Vec.
\end{abstract}

\section{Introduction}

\textcolor{black}{The problem of widespread online misinformation 
has given impetus in recent years to the proliferation of research works that have attempted to curb this phenomenon from different perspectives and with respect to different domains. Distinct approaches have 
been applied mainly 
to review sites, i.e., online platforms that allow users to publish reviews about products and services 
\cite{patel2018survey}, and microblogging platforms, which often disseminate newsworthy content related to politics and events \cite{zhou2020survey}.} 
\textcolor{black}{The majority of these solutions are often based on the identification of particular characteristics (i.e., \emph{features}) that are highly domain-specific, related to the content being disseminated, the purpose of the dissemination, the platform being considered, the authors of the content, and possible 
social interactions in the case of 
social networking sites. Such features are often considered within supervised classifiers categorizing genuine versus non-genuine information, using ``standard'' Machine Learning or more recent deep learning models \cite{viviani2017credibility,girgis2018deep}.}

\textcolor{black}{However, one domain that has been less considered in developing automated solutions for evaluating information credibility, is that of online health-related content dissemination, 
despite the severe harm one might incur in coming into contact with misinformation when searching for possible health treatments and advice. In \cite{chou2018addressing}, \emph{health misinformation} has been defined as: ``a health-related claim of fact that is currently false due to a lack of scientific evidence''. 
In most cases, 
people who are not an expert in the field are unable to properly assess the reliability of such claims, both, in general, due to their limited cognitive abilities \cite{pan2017review} 
and, more specifically, due to their insufficient level of \emph{health literacy} \cite{
sorensen2015health}.} 
%
The difficulties in providing people with automated solutions to compensate for the complexity of evaluating health information on their own, in an online context that is less and less mediated by the presence of medical experts \cite{eysenbach2007intermediation},
lie in the fact that online content related to the health domain
\textcolor{black}{has its own peculiarities compared to other domains of interest. First of all, both long and semi-structured texts published on ``traditional" Web pages (e.g., forums, blogs, question-answering medical systems, etc.), and very short and unstructured texts spread through microblogging platforms (e.g., the mass of COVID-19-related tweets in the last year) are diffused online. Secondly, these texts are characterized by a scientific language and possible reference to external resources that can be taken into account when assessing their credibility.}

\textcolor{black}{With the aim of contributing to social good in the context of studying solutions to prevent people from coming into contact with potentially harmful health misinformation, this work focus on health-related content disseminated in the form of Web pages, an area in which research has 
mainly identified some (handcrafted) features
that make a site or a page ``credible", through the use of 
user-based studies or ML approaches. In this article, the possibility of representing Web pages by means of automatically learned embedding features is explored by considering Web2Vec, a solution recently proposed 
for phishing Web page detection \cite{feng2020Web2vec}. \textcolor{black}{With respect to Web2Vec, we inject some credibility aspects in the feature extraction phase and in the deep learning architecture employed.} Evaluations are performed against publicly available datasets containing health-related content in the form of Web pages, and some baselines that have been proposed in the literature to assess the credibility of the information in both the general and health fields.}



\section{Related Work}

Given the purpose of this article, in this section we focus primarily on describing approaches that have considered the problem of assessing the credibility of health-related Web pages, while mentioning, however, some recent research directions with respect to social platforms.

\subsection{Interactive-based Approaches}\label{IB}

\textcolor{black}{Initially, so-called \emph{interactive-based} approaches applied to Web sites/pages have been considered to assess the credibility (and/or other related concepts partially overlapping or tangent to that of credibility) of health-related information, by employing 
%
interview questionnaires or other interactions with the users.} 
\textcolor{black}{Such approaches are based on users' \emph{subjective} perceptions, which are driven by their information needs and other personal/demographic factors} \cite{diviani2016exploring,song2019role}. 
\textcolor{black}{In} \cite{sbaffi2017trust,kim2016trust}, the authors stated that users’ 
evaluations \textcolor{black}{of Web content} are influenced, \textcolor{black}{in particular,} by \emph{source-related}, \emph{content-related}, and \emph{design-related factors}. 
%

Such factors \textcolor{black}{can have either positive or negative effects on the users' evaluations with respect to the credibility of information.} 
For example, \textcolor{black}{source-related factors such as the authority of} owners/sponsors 
\textcolor{black}{have been shown to have a positive effect on the perception of credibility 
\cite{marton2010women}, even if, in some cases, users tend not to trust communication that is too ``institutional" \cite{williams2003health}, perceived in particular by younger users as  
old and ``not cool" \cite{payton2014online}.} Whereas for content-related factors, the consensus among sources 
has been considered as \textcolor{black}{a good credibility indicator}. Users \textcolor{black}{have shown} mixed feeling towards personal experience and facts. Some users evaluate \textcolor{black}{the presence of} ``objective" facts \textcolor{black}{in a positive way} \cite{williams2003health}, while others find imbalance in their presence \cite{scantlebury2017experiences}. \textcolor{black}{Design-related factors can be evaluated either positively or negatively: in general, an aesthetically well-maintained and easy-to-navigate site is perceived as credible. On the contrary, it is perceived negatively \cite{metzger2003credibility}.}

\textcolor{black}{However,} the 
\textcolor{black}{credibility} of online health\textcolor{black}{-related} information is a complex concept involving more than two dozen dimensions \textcolor{black}{subjectively assessed} by users. \textcolor{black}{For this reason, in recent years, some approaches have been proposed to automatically address the problem both with respect to Web and social media content.} 

\subsection{Automated Approaches}



\textcolor{black}{In \cite{Xie2011}, \textcolor{black}{the authors} presented a framework for predicting \textcolor{black}{the so-called \emph{resource quality} (RQ) of medical Web pages,} by} using \textcolor{black}{an SVM model} trained on 750 resources published on Breast Cancer Knowledge Online (BCKOnline) \cite{malhotra2003breast}. 
The authors described RQ as a 
composition of \emph{reliability} and \emph{relevance}, \textcolor{black}{where} reliability \textcolor{black}{is assessed based on} quality dimensions such as \emph{accuracy}, \emph{credibility}, and \emph{currency} \cite{xie2009sustaining}, \textcolor{black}{and relevance is related to the utility of the page for a user searching for a given medical information.} 
\textcolor{black}{The considered} features \textcolor{black}{are} \textcolor{black}{\emph{categorical features}, i.e., related to the audience (e.g., age, disease stage, etc.), the type (e.g., medical, supportive, and personal), and the subject of the Web page (e.g. treatment-related, therapies-related, etc.)}, 
and \textcolor{black}{simple \emph{textual features}, such as title, description, creator, publisher, and access right}. 

In \cite{sondhi2012reliability}, an automatic approach \textcolor{black}{based on SVM} for \emph{reliability} prediction of medical Web pages 
has been proposed. 
In the approach, the reliability of a Web page is \textcolor{black}{assessed by performing binary classification.} 
The author explored \textcolor{black}{the usage of} \emph{link-based}, \emph{commercial}, \emph{PageRank}, \emph{presentation}, and \textcolor{black}{\emph{textual features}}. \textcolor{black}{Link-based features are related to some counting of internal and external links (and related properties) in the Web page.} 
Commercial features refer to \textcolor{black}{the presence of} commercial \textcolor{black}{terms in the page.} 
PageRank \textcolor{black}{features are related to} 
the relative importance of a Web page computed via PageRank.
\footnote{\url{https://metacpan.org/release/WWW-Google-PageRank/}} 
\textcolor{black}{Presentation features refer} \textcolor{black}{to the clearly in the presentation of content on the page.} 
\textcolor{black}{Finally,} textual features are simply defined as normalised word frequency \textcolor{black}{vectors}. 

\textcolor{black}{Two other recent works based on the use of handcrafted features and Machine Learning approaches are those described in \cite{meppelink2020reliable,fernandez2021reliability}. In
%
\cite{meppelink2020reliable}, a Logistic Regression model for assessing the \emph{reliability} of Web pages has been trained on labeled data collected w.r.t. 
13 vaccine-related search queries. \emph{Textual features} are employed in the form of count-based and TF-IDF word vectors. 
In \cite{fernandez2021reliability}, a replicability study has been conducted on \cite{sondhi2012reliability}, 
considering two additional datasets made available in \cite{schwarz2011augmenting,suominen2018overview}, and 
ignoring PageRank features, deemed as not suitable for assessing Web content reliability \cite{popat2016credibility}}. 

\textcolor{black}{Solutions that attempt to refer to criteria of reliability of medical information provided by external bodies are those proposed by \cite{Boyer2015,Al-Jefri2017,kinkead2020autodiscern,Cui2020}.}
In \cite{Boyer2015}, 
the capability of an automated system to perform the task of identifying 8 HONcode principles on health Web sites has been studied.\footnote{\textcolor{black}{The HONcode certification is an ethical standard aimed at offering quality health information. \url{https://www.hon.ch/cgi-bin/HONcode/principles.pl?English}}} 
Distinct \textcolor{black}{Naive Bayes} classifiers are trained \textcolor{black}{over} \textcolor{black}{a collection of Web pages labeled w.r.t. the considered criteria. In this approach,} 
Web site content is converted into \textcolor{black}{weighted bag-of-words representations.} 
In \cite{Al-Jefri2017}, 
to confirm the \emph{evidence-based medicine} (EBM) \textcolor{black}{property} of a Web page, the treatment \textcolor{black}{described} in the Web page is checked \textcolor{black}{w.r.t. its approval} by the US Food and Drug Administration, \textcolor{black}{the} UK National Health System, or the National Institute of Care Excellence. Two different \textcolor{black}{feature types} are \textcolor{black}{considered for classification:} \emph{text-based} and \emph{domain-specific} features, \textcolor{black}{related to} JAMA criteria \cite{silberg1997assessing}. 
A number of distinct ML classifiers have been used for the experiments. 
\textcolor{black}{In \cite{kinkead2020autodiscern}, the authors have proposed to automate the use of DISCERN.\footnote{DISCERN is a questionnaire 
providing users with a way of assessing the quality of information on health treatment choices. \url{http://www.discern.org.uk/}} Five \emph{Hierarchical Encoder Attention-based} (HEA) models (related to 5 DISCERN criteria) are trained on articles related to breast cancer, arthritis, and depression. 
A \emph{Bidirectional Recurrent Neural Network} (BRNN) layer converts words, sentences, and documents to dense vector representations. \textcolor{black}{Such representations are used for classification} using a softmax layer.
}
A knowledge-guided graph attention network named DETERRENT for detecting \emph{health misinformation} has been recently proposed in \cite{Cui2020}, trained on articles related to diabetes and cancer. It incorporates a Medical Knowledge Graph and an Article-Entity Bipartite Graph, and propagates node embeddings \textcolor{black}{representing Web pages} through Knowledge Paths 
for misinformation classification.

\textcolor{black}{
More recently, the interest of the scientific community to detect health misinformation is turning to the use of 
\emph{deep learning} solutions, especially w.r.t. social media content \cite{Samuel2018,bal2020analysing}, 
achieving promising results.} However, w.r.t. the tangent problem of phishing Web page detection, a recent deep learning approach presented in 
\cite{feng2020Web2vec}, i.e., Web2Vec, has been proposed. 
\textcolor{black}{Hence, 
in this paper we intend to study such a solution (suitably modified) with respect to the problem of assessing the credibility of Web page content, by considering, in the deep learning model, source, content, and design factors, together with contextual aspects related to the presence of links and medical-related terms in the Web page.}


\section{A Web2Vec-based Solution for Health Misinformation Detection}
\label{sec:pm}


The original Web2Vec model \cite{feng2020Web2vec}, developed for phishing Web page detection, is based on the 
embedded representation of the URL, content, and DOM structure of the considered Web page. Such embedding representations are used by a hybrid CNN-BiLSTM network to extract local and global features, which are combined by an attention mechanism strengthening important features. Multi-channel output vectors are concatenated and provided to a classifier to determining the category of the tested Web page (i.e., phishing vs non-phishing).

\textcolor{black}{In the approach proposed in this article, in the application and in the appropriate modification of the Web2Vec model, some characteristics related to the problem of assessing the credibility of health information are taken into account. First of all, when generating an embedded representation of Web pages, the use of a specific vocabulary related to the medical field is considered, which is crucial to detect health misinformation. In addition, instead of focusing on the features related to the URL of the Web page to be evaluated (as Web2Vec did), those related to the URLs present in the page itself are considered, because they can give a better indication of whether they refer to reliable or unreliable external sources (e.g., the presence of commercial links). With these aspects in mind, the proposed solution consists of the following phases:}
\begin{itemize}
    \item \textit{Data Parsing}: page links, content, and Document Object Mode (DOM) structure are parsed from each HTML page in the dataset, \textcolor{black}{to extract suitable data that are employed in the following phase};
    \item \textit{Data Representation}: word-level and sentence-level embedding representations are generated for the Web page content, while for the DOM structure and links, HTML tags and URLs embeddings are considered;
    \item \textit{Feature Extraction}: a CNN-BiLSTM network is used to extract features from the given representations;
    \item \textit{Web Page Classification}: health-related Web pages are classified as credible or not credible by using densely connected layers.
\end{itemize}

\subsection{Data Parsing} \label{sec:dp}

\textcolor{black}{The data parsing operation is the same as in Web2Vec, with the exception of link parsing, which in the case of this approach is applied to the content of the HTML page.}

\paragraph{DOM Corpus.}\label{dom}

HTML files are characterized by a typical semi-structured data format. The hierarchical structure 
is represented using HTML tags, \textcolor{black}{organized according to the Document Object Model (DOM) structure. Focusing on such a structure, we extracted an ordered list of tags, starting from high-level tags} 
until ``children'' tags, i.e., \texttt{HTML}, \texttt{HEAD}, \texttt{META}, \texttt{LINK}, \texttt{TITLE}, \texttt{SCRIPT}, \texttt{BODY}, \texttt{DIV}, \texttt{TABLE}, \texttt{TR}, \texttt{TD}, \texttt{IMG}. 
%
%
Such HTML tags are considered as words, \textcolor{black}{which constitute the \emph{world-level corpus} for the DOM structure to be used in the data representation phase.}

\paragraph{Content Corpus.} \label{pcc}

Each Web page is phrased, and only \textcolor{black}{textual content is considered (links and tags are excluded). Both a \emph{word-level} and a \emph{sentence-level} corpus are constructed. The first is constituted by each distinct word present in the page, the second identifies} word sequences separated by the ` . ' character. 
Specifically, we consider a fixed-length dimension for each word sequence. 


\paragraph{Link Corpus.}

\begin{figure}[ht]
\centering
\includegraphics[width=0.8\columnwidth]{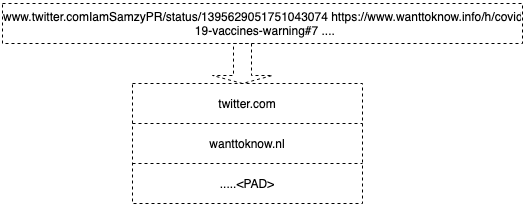}
\caption{\textcolor{black}{Construction of the word-level corpus for links.}}
\label{fig:url}
\end{figure}

The link corpus is created considering links present in the HTML page. In particular, we focus on the domain names \textcolor{black}{extracted from the} URL of the Web sites referenced in the HTML page. 
\textcolor{black}{Such domain names, illustrated in Figure \ref{fig:url}, represent the} \emph{word-level corpus} \textcolor{black}{to be employed in the data representation phase.}

\subsection{Data Representation}

\textcolor{black}{In this phase, the word- and sentence-level corpora related to Web page DOM structure, content, and links, generated in the previous phase, are formally represented in order to capture their semantic relationships through word embedding. In particular,}
%
%
a Keras embedding layer is employed,\footnote{\url{https://keras.io/api/layers/core_layers/embedding/}} \textcolor{black}{which is based on a supervised method that enhance the semantic representation while training the model using backpropogation. 
It is worth to be underlined that a separate embedding layer is defined for the DOM corpus, the word-level content and link corpora, and the sentence-level content corpus.} 

\begin{figure}[ht]
\centering
\includegraphics[width=0.8\columnwidth]{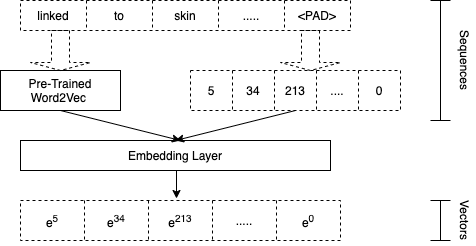}
\caption{The word-level embedding phase for the content.}
\label{fig:Embed}
\end{figure}

\textcolor{black}{With respect to Web2Vec, in this work,} to include domain-specific information related to the medical field, we 
add 
a word2vec layer pre-trained on PubMed as a weight initializer in the Keras embedding layer when considering the content word-level embedding. 
%
\textcolor{black}{In this way, t}he word2vec weights are used as weight initializers for the embedding layer, as illustrated in Figure \ref{fig:Embed}. 



\subsection{Feature Extraction}

\textcolor{black}{The features, as in the case of Web2Vec, are extracted by means of a CNN-BiLSTM network with an attention mechanism applied to the embedding representations obtained in the previous phase. \emph{Convolutional Neural Networks} (CNN)s are}
%
%
nowadays commonly used for local feature extraction from data. 
\emph{Bidirectional Long Short-Term Memory} (Bi-LSTM) networks are used to overcome the ability to learn feature from sequences \textcolor{black}{of CNN by combining the word with its context} \cite{fan2018neural}. The attention mechanism is used to improve the prediction capacity of the model.

\paragraph{CNN.}

The employed CNN is constituted, \textcolor{black}{as in Web2Vec,} by a feed-forward network model structure. The hidden layer is divided into a convolution layer and a pooling layer. 
To overcome over-fitting, each fully connected layer is followed by one dropout layer (with a dropout ratio of 0.05). \textcolor{black}{Details on the convolution and pooling operations can be found in \cite{feng2020Web2vec}}.
%
%
%
%
%

\paragraph{BiLSTM.}

The output of the CNN 
layer constitutes the input of the BiLSTM layer. 
Such layer is formed using  Long Short-Term Memory in both directions, i.e., forward and backward, which keeps the sequential order among the data. It also allows detecting the relationship between the previous inputs and the output. 

Since BiLSTM is a sequential- and memory-based model, it can both learn long-term dependence on the Web page and also extract improved features using local features from CNN. 
To deal with possible overfitting, \emph{dropout learning} and \emph{L2 regularization} (as detailed in the next section) are used to improve the model training.

\paragraph{Attention Layer.}

\textcolor{black}{The addition of the attention layer, in the case of assessing the credibility of health-related information, is dictated by the fact that in the same document there may be parts characterized by ``more credible" and ``less credible" information. In this situation, even the presence of a small amount of ``non-credible" features characterizing a credible page (or vice versa), can negatively affect its final evaluation}. The purpose of the attention layer is, therefore, to pay particular attention with respect to the most discriminant features w.r.t. the considered problem; in this work, we have referred in particular to the concept of \emph{additive attention} \cite{bahdanau2014neural}. 


\subsection{Web Page Classification}

\textcolor{black}{Web pages are categorized into  credible and not credible through the use of a \emph{binary classifier} consisting of} a fully connected layer \textcolor{black}{having a} sigmoid function in the final layer, which combines the features extracted from the previous layers relating to the four corpora considered (DOM corpus, word-level content and link corpora, sentence-level content corpus).

For the classification loss calculation, the \emph{cross-entropy loss function} and the \emph{L2 regularization} are applied to overcome overfitting. 
\textcolor{black}{Formally:} 
%
\[
Error(t - y) = - \frac{1}{N} \sum_n^N
  \left[ t_n \ln y_n + (1-t_n) \ln (1-y_n) \right]
\]
and
%
\[Loss = Error(t - y) +  \lambda \sum_{n}^N w_n^{2}\]
where $t$ is the target label, $y$ the predicted label, $w$ the weight matrix of the layer, and $\lambda$ is the so-called \emph{L2 penalty parameter}.


\section{Experimental Evaluations}

In this section, we present the experimental evaluation of the effectiveness of the proposed model. Specifically, we introduce the description of the different datasets, baselines, and evaluation metrics considered, together with technical and experimental details followed by a discussion on the obtained results.

\subsection{Description of the Datasets}
\label{data}

Only a few publicly available datasets are currently available for evaluating health-related information w.r.t. credibility. In particular, it has been necessary to consider those datasets from which it was possible to obtain the original HTML format of Web pages. Hence, the choice has fallen on the datasets provided by 
\cite{sondhi2012reliability,schwarz2011augmenting,goeuriot2020overview}.

\paragraph{Microsoft Credibility Dataset \cite{schwarz2011augmenting}.}
This dataset is constituted by 1,000 Web pages in different domains such as Health, Finance, Politics, etc. 
Credibility ratings associated with them are provided over a five-point Likert scale, ranging from 1 to 5, where 1 stands for “very non-credible”, and 5 for “very credible". 
\textcolor{black}{In \cite{fernandez2021reliability}, for evaluation purposes,} labels have been pre-processed by removing the middle value 3, and mapping 4-5 rating values to credible Web pages and 1-2 rating values to non-credible Web pages. \textcolor{black}{In our approach, we followed the same strategy, and} we focused on the 130 available health-related Web pages. 
Out of 130, 104 were credible and 26 were non-credible. Given the high data imbalance, we applied the SMOTE \cite{chawla2002smote} oversampling method to the minority class.

\paragraph{Medical Web Reliability Corpus \cite{sondhi2012reliability}.}

This is a manually generated balanced dataset with binary (i.e., reliable and unreliable) labels associated with Web pages. The authors randomly selected reliable Web sites from HON accredited Web sites.\footnote{\url{https://www.hon.ch/en/}} Unreliable Web sites were searched on the Web using queries, constituted by the disease name + ``miracle cure". 
The dataset consists of 360 Web pages, 180 reliable and 180 unreliable. \textcolor{black}{After a cleaning phase, to remove blank and no-longer accessible pages,}
%
we dealt with 170 reliable Web pages and 176 unreliable Web pages.

\paragraph{CLEF eHealth 2020 Task-2 Dataset \cite{goeuriot2020overview}.}

\textcolor{black}{This dataset consists of a larger number of documents than the previously illustrated datasets, and has been expressly built to assess the topical relevance, readability, and credibility of Web pages consisting of medical content, as part of the so-called \emph{Consumer Health Search} (CHS) task.\footnote{\url{https://clefehealth.imag.fr/?page_id=610}} Credibility ratings are expressed on a}
four-point scale, from 0 to 3.
%
Such ratings have been converted to binary values by considering 0-1 values as  non-credible and 1-2 values as credible. 
Finally, we dealt with 5,509 credible and 6,736 non-credible Web pages.



\subsection{Baselines and Evaluation Metrics}\label{sec:baselinesevaluation}

The approaches that are taken into consideration as baselines for assessing the effectiveness of the proposed approach concern solutions developed for assessing the credibility of both ``general" and health-related information, \textcolor{black}{which consider both textual and other families of handcrafted features in association with Machine Learning.} In particular, we consider the textual-feature-based model proposed in \cite{meppelink2020reliable}, the multi-feature-based model proposed in \cite{fernandez2021reliability}, which encompasses another multi-feature-based model discussed in \cite{sondhi2012reliability}, and a BioBERT-SVM model \textcolor{black}{that has been developed in this work for evaluation purposes, given that BERT embeddings have produced a good result in association with SVM in fake news and misinformation detection problems \cite{glazkova2020g2tmn, dharawat2020drink, karande2021stance}. Specifically, with respect to this baseline, we consider BERT embeddings pre-trained on PubMed articles for adapting to the biomedical domain \cite{lee2020biobert}.}




With respect to the above-mentioned baselines, the following evaluation metrics are taken into consideration: F1 measure, accuracy and AUC. Such metrics have often been used in various literature works related to misinformation detection and credibility assessment \cite{Cui2020, meppelink2020reliable}. For training the ML models employed as baselines, the \texttt{scikit-learn} library \cite{sklearn_api} has been used.\footnote{\url{https://scikit-learn.org/}} To evaluate the results, 5-fold stratified cross-validation has been applied. \textcolor{red}{}

\subsection{Results and Discussion}

This section illustrates and discusses the 
results of the proposed solution with respect to each dataset and baseline described in the previous sections, in terms of the 
above-mentioned evaluation metrics respectively. 
In the following, the considered \emph{baselines} are denoted as: NB-CountVec and LR-TF-IDF, identifying the most effective approaches presented in \cite{meppelink2020reliable}, based on the application of a Naive Bayes and a Logistic Regression classifier to textual features expressed as count vectors and TF-IDF vectors; MFB-SVM, denoting the multi-feature model presented in \cite{fernandez2021reliability}; and BioBERT-SVM, as detailed in Section \ref{sec:baselinesevaluation}.
%
Furthermore, two Web2Vec variations have been considered:\footnote{It was not possible to evaluate Web2Vec also w.r.t. its original link embedding model because in the considered datasets the URLs of the considered pages were missing.} 
\begin{itemize}
\item Web2Vec(C): it refers to the Web2Vec model trained only on content embeddings with default weight initializers;
\item Web2Vec(C-D): it refers to the Web2Vec model trained on both content and DOM embeddings with default weight initializers.
\end{itemize}

Such additional baselines have been compared with distinct instantiations of the proposed model based on Web2Vec for assessing credibility, denoted as Cred-W2V. In particular:

\begin{itemize}
\item Cred-W2V(C): it refers to the proposed model trained on content embeddings with the PubMed word2vec layer acting at weight initializer;
\item Cred-W2V(C-D): it refers to the proposed model trained on content embeddings with the PubMed word2vec layer, and on DOM embeddings with default weights;
\item Cred-W2V(C-D-L): it refers to the proposed model trained on content, DOM, and link embeddings with default weight initializers;\footnote{With respect to the original Web2Vec model, we recall that links are referred to those URLs present in a Web page.}
\item Cred-W2V(C-D-L)*: it refers to the proposed model trained on content, DOM, and link embeddings, with the PubMed word2vec layer acting at weight initializer.
\end{itemize}

The considered datasets (Section \ref{data}), are denoted as D1 (Microsoft Credibility Dataset), D2 (Medical Web Reliability Corpus), and D3 (CLEF eHealth 2020 Task-2 Dataset). \textcolor{black}{Only for D3, it was possible to calculate, given the higher number of labeled data, the \emph{binomial proportion confidence intervals} with 95\% confidence 
\cite{blyth1983binomial}. In this case, the results are reported under the label D3(BI).}


\begin{table}[!htp]\centering
\caption{Evaluation results.}\label{table:2}
\small
\begin{tabular}{lrrrrr}\toprule
&Metrics &D1 &D2 &D3 &D3(BI) \\\midrule
\multirow{3}{*}{NB-CountVec} &Accuracy &74.55 &94.43 &64.89 &$64.9\pm3.00$ \\
&F1 &83.22 &94.71 &67.84 &$67.2\pm3.00$\\
&AUC &67.02 &93.98 &64.12 &$64.6\pm2.93$\\ \hline
\multirow{3}{*}{LR-TF-IDF} &Accuracy &75.35 &94.29 &68.6 &$67.9\pm2.55$ \\
&F1 &85.82 &94.37 &71.3 &$70.9\pm2.80$ \\
&AUC &47.18 &93.21 &67.6 &$67.8\pm2.55$ \\ \hline
\multirow{3}{*}{MFB-SVM} &Accuracy &70.03 &94.73 &66.15 &$63.8\pm3.50$ \\
&F1 &75.97 &93.52 &46.03 &$46.7\pm2.50$ \\
&AUC &57.44 &93.98 &47.78 &$50.2\pm0.10$ \\ \hline
\multirow{3}{*}{BioBERT-SVM} &Accuracy &72.1 &94.1 &70.74 &$69.8\pm2.00$\\
&F1 &44.67 &94.2 &65.34 &$65.3\pm4.00$ \\
&AUC &63.2 &94.1 &69.56 &$67.0\pm3.00$\\ \hline
\multirow{3}{*}{Web2Vec(C)} &Accuracy &\textit{78.34} &\textit{94.81} &70.34 &$69.5\pm2.50$\\
&F1 &\textit{85.67} &\textit{94.49} &\textit{71.56} &$\mathit{68.9\pm2.75}$\\
&AUC &\textit{65.34} &\textit{94.54} &\textit{70.18} & $\mathit{68.9\pm2.10}$ \\ \hline
\multirow{3}{*}{Cred-W2V(C)} &Accuracy &\textit{78.34} &\textbf{\textit{96.1}} &\textbf{\textit{71.38}} &\bm{$\mathit{71.5\pm1.75}$} \\
&F1 &\textbf{\textit{86.34}} &\textbf{\textit{95.21}} &\textbf{\textit{72.35}} &\bm{$\mathit{71.8\pm2.25}$} \\
&AUC &\textbf{\textit{68.13}} &\textbf{\textit{95.98}} &\textbf{\textit{71.59}} &\bm{$\mathit{70.9\pm2.10}$} \\ \hline
\multirow{3}{*}{Web2Vec(C-D)} &Accuracy &\textit{80.7} &\textit{96.4} &\textit{72.12} &$\mathit{71.9\pm2.22}$ \\
&F1 &\textit{88.28} &\textit{96.12} &\textit{73.69} &$\mathit{72.5\pm1.70}$ \\
&AUC &\textit{74.34} &\textit{96.32} &\textit{71.71} &$\mathit{71.1\pm1.75}$\\ \hline
\multirow{3}{*}{Cred-W2V(C-D)} &Accuracy &\textbf{\textit{86.9}} &\textbf{\textit{97.57}} &\textbf{\textit{73.58}} &\bm{$\mathit{72.5\pm2.20}$} \\
&F1 &\textbf{\textit{91.62}} &\textbf{\textit{97.69}} &\textbf{\textit{77.98}} &\bm{$\mathit{75.5\pm2.15}$} \\
&AUC &\textbf{\textit{80.07}} &\textbf{\textit{97.42}} &\textbf{\textit{73.59}} &\bm{$\mathit{72.8\pm1.40}$} \\ \hline
\multirow{3}{*}{Cred-W2V(C-D-L)} &Accuracy &\textbf{\textit{84.12}} &\textbf{\textit{96.23}} &\textbf{\textit{73.98}} &\bm{$\mathit{73.4\pm1.70}$} \\
&F1 &\textbf{\textit{90.45}} &\textbf{\textit{96.24}} &\textbf{\textit{75.74}} &\bm{$\mathit{75.1\pm1.97}$}\\
&AUC &\textbf{\textit{78.17}} &\textbf{\textit{96.26}} &\textbf{\textit{73.85}} &\bm{$\mathit{73.4\pm2.10}$}\\ \hline
\multirow{3}{*}{Cred-W2V(C-D-L)*} &Accuracy &\textbf{\textit{90.75}} &\textbf{\textit{98.81}} &\textbf{\textit{74.42}} &\bm{$\mathit{75.5\pm2.35}$}\\
&F1 &\textbf{\textit{94.17}} &\textbf{\textit{97.77}} &\textbf{\textit{76.62}} &\bm{$\mathit{78.5\pm2.15}$} \\
&AUC &\textbf{\textit{86.17}} &\textbf{\textit{97.68}} &\textbf{\textit{74.24}} &\bm{$\mathit{75.8\pm1.50}$} \\ \hline
\bottomrule
\end{tabular}
\end{table}


\textcolor{black}{As it can be seen from Table \ref{table:2}, the proposed model for health misinformation detection outperforms all the baselines that rely on the use of handcrafted features and Machine Learning techniques with respect to all datasets and evaluation metrics considered (values in ``italic"). Also compared to using the original Web2Vec model applied to the problem considered in this article, the proposed model allows to obtain better}
results (values in ``bold"), both when the word2vec layer trained on PubMed is added to the original architecture, and when we consider the embeddings of the links present in the pages to be evaluated. 
%
%
\textcolor{black}{We can say in particular, looking at the comparison of the results of the Cred-W2V(C-D), Cred-W2V(C-D-L) and Cred-W2V(C-D-L)* models, that the impact of including a pre-trained embedded representation on a domain-specific lexicon is preponderant over the effectiveness of the proposed approach. However, this may be due to the fact that, in the considered datasets, some HTML pages presented no internal links, in some cases due to the original gathering process itself.}

\section{Conclusions}


In this article, we addressed the issue of identifying misinformation in health-related Web pages. 
Starting from an analysis of the literature works, 
we identified some of the most frequently adopted features and solutions 
to address the considered problem. In particular, ``classical" Machine Learning techniques and a deep learning approach recently proposed for detecting phishing Web pages, i.e., Web2Vec.
Starting from this model, we have applied and suitably modified it w.r.t. the problem of health misinformation detection, concluding that such an approach can be effective when customized on the considered domain and, in any case, it is more effective than traditional ML-based methods. 

This work represents a first step for the investigation at a more general level of what can be actually the most effective architectures and features to solve the problem; from the work it emerged that actually a semantic and context-aware representation of the text contained in the Web pages has a big impact on the effectiveness of the identification of health misinformation. However, the distinct impact of the presence of links and structural features of the page is certainly worth investigating in the future, together with
the possibility to insert into the model domain knowledge also through the addition of both handcrafted and other embedding features.

\section*{Acknowledgements}

This work is supported by the EU Horizon 2020 ITN/ETN on Domain Specific Systems for Information Extraction and Retrieval (H2020-EU.1.3.1., ID: 860721).




\bibliographystyle{plain}
\bibliography{hmd}


\end{document}